\documentclass[11pt]{article}
\usepackage[english]{babel}
\usepackage{amsmath,amssymb,amsthm}

\usepackage[pdfpagemode=UseOutlines,
bookmarks=true, bookmarksopenlevel=3, bookmarksopen=true,
bookmarksnumbered=true, unicode=true,
pdffitwindow=true,pdfpagelayout=SinglePage,pdfhighlight=/P,
colorlinks=true,linkcolor=blue,citecolor=blue,urlcolor=blue]{hyperref}

\usepackage[left=2.5cm, right=2.5cm, top=2.5cm, bottom=2.5cm]{geometry}
\allowdisplaybreaks[4]

\theoremstyle{plain}
\newtheorem{proposition}{Proposition}
\theoremstyle{remark}
\newtheorem{remark}[proposition]{Remark}

\title{Negative flows for several integrable models}

\author{V.E.\:Adler\thanks{L.D.\:Landau Institute for Theoretical Physics, Akad. Semenova av. 1A, 142432, Chernogolovka, Russian Federation. E-mail: adler@itp.ac.ru}}

\date{October 17, 2023}

\begin{document}
\maketitle

\begin{abstract}
A construction of negative flows for integrable systems based on the Lax representation and squared eigenfunctions is proposed. Examples considered include the Boussinesq equation and its reduction to the Sawada--Kotera and Kaup--Kupershmidt equations; one of the Drinfeld--Sokolov systems and its reduction to the Krichever--Novikov equation.
\end{abstract}

\section{Introduction}

A characteristic feature of equations integrable by the inverse scattering method is the existence of an infinite hierarchy of evolutionary higher symmetries. However, the matter is not limited to them: in addition, there are so-called negative symmetries, written in the form of hyperbolic equations or in the form of evolution equations with non-local variables. Often such symmetries are of interest in themselves, for example, the negative flow for the Korteweg--de Vries equation (KdV) associated with the Camassa--Holm equation \cite{Schiff_1998, Hone_1999}, sine-Gordon type equations that define hyperbolic symmetries for equations of the KdV type \cite {Meshkov_Sokolov_2011}, the Maxwell--Bloch system, which determines the negative symmetry of the nonlinear Schr\"odinger equation (NLS) \cite{Rogers_Schief_2002} and many other examples.

One of possible definitions of negative flows can be given based on recursion operators. Let an integrable hierarchy be generated according to the formula $u_{t_n}=R^{n-1}(f)$ where $R$ is the recursion operator and $u_{t_1}=f[u]$ is a seed symmetry. Then the negative symmetry can be defined by inverting the operator $R-\alpha$ with arbitrary constant $\alpha$ (which is natural since this is also a recursion operator), that is, as the the flow of the form
\[
 u_z= (R-\alpha)^{-1}(f)\quad\Leftrightarrow\quad (R-\alpha)(u_z)=f.
\]
Moreover, in most examples, $R$ contains integral terms of the form $fD^{-1}\rho$ (we will use the notations $D=\partial_x$ and $D^{-1}$ for the derivative and the antiderivative with respect to the spatial variable $x$) and then the result of applying $R$ to $u_z$ includes the term $cf$ with an arbitrary integration constant. By choosing $c=1$, we can reduce the previous definition to the form
\[
 R(u_z)=\alpha u_z,
\]
that is, the negative flow serves as an eigenfunction of the recursion operator. The expansion $(R-\alpha)^{-1}=-\alpha^{-1}(1+R/\alpha+R^2/\alpha^2+\dots)$ suggests that this flow can be viewed as the generating function for the usual higher symmetries. 

This approach works very well if $R$ is of low order. An example of its implementation for the NLS equation can be found in \cite{Kamchatnov_Pavlov_2002}, see also \cite{Aratyn_Gomes_Zimerman_2006}; an example for the Volterra lattice can be found in \cite{Adler_2023}. However, there are many equations for which $R$ is rather complicated. In this paper we are trying to answer the question of how to derive the equations of negative symmetry in such a situation and whether they can be somehow simplified. As examples, we take the hierarchies of the Boussinesq equation \cite{Zakharov_Shabat_1974, Ablowitz_Haberman_1975} and one of the Drinfeld--Sokolov systems (DS-III) \cite{Drinfeld_Sokolov_1985}. In both cases, we start from the known Lax representation of the hierarchy
\[
 L_{t_n}=[A_n,L]
\]
where $L$ and $A_n$ are differential operators. The recursion operator $R$ for these hierarchies is also known and this allows us to immediately write down the negative symmetry, which turns out to be rather cumbersome. To simplify it, we first analyze the scheme for deriving $R$ from the Lax representation proposed in the paper \cite{Gurses_Karasu_Sokolov_1999} (which seems to provide the most direct method) and demonstrate that negative symmetry also admits a Lax representation of the form
\begin{equation}\label{i.negLax}
 L_z(L-\alpha)=[P,L] \quad\Leftrightarrow\quad L_z=[P(L-\alpha)^{-1},L] 
\end{equation}
where $P$ is a differential operator of order less than the order of $L$. By use of this equation, negative symmetry is calculated more easily and without explicitly using $R$. Moreover, although this does not change the answer, it does suggest a way for further simplification: it turns out that the nonlocal variables involved in negative symmetry can be represented by the product and the Wronskian of eigenfunctions of the operator $L$. This establishes a connection between the negative symmetry and the method of squared eigenfunctions (see, for example \cite{Fokas_Anderson_1982}). In new variables, the order of the equations decreases.

The simplest illustration of this scheme is the KdV equation
\[
 u_t=u_{xxx}-6uu_x,
\]
which corresponds to the operator $L=-D^2+u$. Here the operator $P$ is of the form $P=-\frac{1}{2}gD+\frac{1}{4}g_x$ and equation (\ref{i.negLax}) amounts to relations
\[
 u_z=g_x,\quad g_{xxx}-4(u-\alpha)g_x-2u_xg=0.
\]
The second equation is equivalent to $R(g)=-4\alpha g$ where $R=D^2-4u-2u_xD^{-1}$ is the well-known recursion operator for KdV, and it allows for reduction of order by integration with the factor $2g$:
\begin{equation}\label{KdVneg}
 u_z=g_x,\quad 2gg_{xx}-g^2_x-4(u-\alpha)g^2+w^2=0,\quad w_x=0.
\end{equation}
These equations determine the negative KdV symmetry, also known as the associated Camassa--Holm equation \cite{Schiff_1998, Hone_1999}. In this example, the integrating factor is easy to guess, but there is also a more systematic way to reduce the order, by passing to variables $g=\varphi\psi$ and $w=\varphi\psi_x-\varphi_x\psi$ where $\psi$ and $\varphi$ are eigenfunctions of $L$ (in general, this is not equivalent to the use of first integrals).

This scheme is described in more detail in section \ref{s:Bsq} for the example of the Boussinesq hierarchy associated with the third-order operator $L$. Note that similar results were obtained quite recently in paper \cite{Lou_Jia_2023}. We consider also reductions to the Sawada--Kotera \cite{Sawada_Kotera_1974} and Kaup--Kupershmidt equations \cite{Kaup_1980}. For them, among the general negative symmetries, one can distinguish simpler degenerate negative symmetries corresponding to the value $\alpha=0$ and associated with the Degasperis--Procesi and Tzitzeica equations \cite{Degasperis_Procesi_1999, Degasperis_Hone_Holm_2002}.

Section \ref{s:DS3} is devoted to the Drinfeld--Sokolov system and its reduction to the Krichever--Novikov equation \cite{Krichever_Novikov_1980}. For this system, $L$ is the self-adjoint fourth-order operator and all the formulas turn out to be a little more complicated, but the general scheme remains practically unchanged and we present only the key formulas.

\section{Boussinesq equation}\label{s:Bsq}

\subsection{Lax representations and the recursion operator}\label{s:Bsq.LR}

Our first example is related to the system
\begin{equation}\label{Bsq}
 u_t= -3u_{xx}+6v_x,\quad v_t= -2u_{xxx}+3v_{xx}-2uu_x,
\end{equation}
which is equivalent to the Boussinesq equation
\begin{equation}\label{Bsq.u}
 u_{tt}+3(u_{xx}+2u^2)_{xx}=0
\end{equation}
and admits the Lax representation \cite{Zakharov_Shabat_1974, Ablowitz_Haberman_1975}
\begin{equation}\label{Bsq.LA}
 L_t=[A,L],\quad L=D^3+uD+v,\quad A=3D^2+2u.
\end{equation}
The hierarchy of the system (\ref{Bsq}) is determined by equations
\begin{equation}\label{Bsq.LAn}
 L_{t_n}=[A_n,L],\quad A_n=(L^{n/3})_+,\quad n\in{\mathbb N},~~ n\ne3k
\end{equation}
where $M_+$ denotes the differential part of a pseudo-differential operator $M$. In particular, equations (\ref{Bsq}) correspond to the operator $A=3A_2$, that is, the derivation $\partial_t$ coincides with $3\partial_{t_2}$. 

Let us recall, for this example, the method of deriving the recursion operator from the Lax representation, proposed in the paper \cite{Gurses_Karasu_Sokolov_1999}. The formula $A_n=(L^{n/3})_+$ implies the relations
\[
 A_{n+3}= (L^{n/3}L)_+ = ((A_n+(L^{n/3})_-)L)_+ = A_nL +((L^{n/3})_-L)_+ = A_nL+B_n
\]
where $B_n$ is a differential operator of order not greater than 2. Then
\begin{equation}\label{LAB}
 L_{t_{n+3}}= [A_nL+B_n,L]= A_nL^2-LA_nL+[B_n,L]= L_{t_n}L+[B_n,L],
\end{equation}
where the left-hand side is a first-order operator and the right-hand side is a fourth-order operator. This gives 5 equations; three of them allow us to express all coefficients $B_n$ in terms of coefficients $L$ and $L_{t_n}$. Then
substituting the found expressions into the two remaining equations gives the relationships between the coefficients of $L_{t_{n+3}}$ and $L_{t_n}$, which is written as a mapping $R:\binom{u}{v}_{t_n}\mapsto\binom{u}{v}_{t_{n+3}}$. Calculations according to the described scheme lead to the operator
\begin{equation}\label{Bsq.R}
\begin{gathered}[b]
 -3R=\begin{pmatrix}
  D^3+uD+2u_x-3v & -2D^2-2u\\
  \frac{2}{3}D^4+\frac{4}{3}uD^2+2u_xD+2u_{xx}+\frac{2}{3}u^2-2v_x & -D^3-uD-3v
 \end{pmatrix}\\
 \qquad -\frac{1}{3}\binom{-3u_{xx}+6v_x}{-2u_{xxx}+3v_{xx}-2uu_x}D^{-1}(1,0) 
   -\binom{u_x}{v_x}D^{-1}(0,1)
\end{gathered}
\end{equation}
(this coincides with the operator from \cite{Gurses_Karasu_Sokolov_1999} up to the changes between (\ref{Bsq.u}) and (\ref{Bsq})). The column factors at the integral terms in $R$ correspond to the flows $\partial_t$ and $\partial_x=\partial_{t_1}$. When $R$ acts to a symmetry, $\partial_t$ and $\partial_x$ are added to the answer with arbitrary integration constants. The same two flows play the role of the seed flows: the entire hierarchy consists of the flows $\partial_{t_{3k+1}}=R^k(\partial_{t_1})$ and $\partial_{t_{3k+2}}=R^k(\partial_{t_2})$. According to the explicit formula (\ref{Bsq.LAn}), integral terms in $R$ do not create nonlocalities. Since the coefficients of the integral terms form a non-singular matrix, this means that any flow of the hierarchy can be represented in the form
\begin{equation}\label{Bsq.utau}
 u_\tau=p_x,\quad v_\tau=q_x
\end{equation}
where $p$ and $q$ are local functions of $u$, $v$ and their derivatives.

Let us define the negative flow $\partial_z$ as the generating series for the flows (\ref{Bsq.LAn}):
\begin{equation}\label{Bsq.Dz}
 \partial_z=(c_1\partial_{t_1}+c_2\partial_{t_2}) +\alpha^{-1}(c_1\partial_{t_4}+c_2\partial_{t_5}) +\alpha^{-2}(c_1\partial_{t_7}+c_2\partial_{t_8})+\dots
\end{equation}
where $c_1$ and $c_2$ are arbitrary constants. Since all terms are of the form (\ref{Bsq.utau}), hence $\partial_z$ should be sought in the same form, but, unlike ordinary symmetries, the variables $p$ and $q$ for this flow are not local. They are defined by equation
\begin{equation}\label{Bsq.Ruvz}
 R\binom{u_z}{v_z}=\alpha\binom{u_z}{v_z}
\end{equation}
and straightforward calculations give the following answer. In it, we neglect the integration constants when applying $R$, since they are compensated by adding constants to $p$ and $q$.

\begin{proposition}\label{pr:Bsq.uvzpq}
The negative symmetry for the system (\ref{Bsq}) is determined by equations
\begin{gather}
\label{Bsq.uvz}
 u_z=p_x,\quad v_z=q_x,\\[4pt]
\label{Bsq.pqx}
\left\{\begin{aligned}
  & p_{xxxx}+up_{xx}+2(u_x-3(v-\alpha))p_x+(u_{xx}-2v_x)p-2q_{xxx}-2uq_x-u_xq=0,\\[3pt]
  & 2up_{xxx}+6(v-\alpha)p_{xx}+2(2v_x+u^2)p_x+(v_{xx}+2uu_x)p\\
  & \qquad\qquad +q_{xxxx}+uq_{xx}+3(2u_x-3(v-\alpha))q_x+(2u_{xx}-3v_x)q=0.
\end{aligned}\right.
\end{gather}
\end{proposition}

It turns out that these equations can be derived without explicitly using the recursion operator. To do this, we apply the relations (\ref{LAB}) directly to the flow (\ref{Bsq.Dz}), which gives
\[
 L_z-c_1L_{t_1}-c_2L_{t_2}=\alpha^{-1}L_zL+[\alpha^{-1}(c_1B_1+c_2B_2)+\alpha^{-2}(c_1B_4+c_2B_5)+\dots,L].
\]
Since $L_{t_1}=[A_1,L]$ and $L_{t_2}=[A_2,L]$, we arrive at the operator equation
\begin{equation}\label{LP}
 L_z(L-\alpha)=[P,L],  
\end{equation}
which can be taken as the Lax representation for negative symmetry. As noted in the introduction, this equation can be brought to the usual Lax form, but with a non-standard operator $A$:
\begin{equation}\label{LP'}
 L_z=[P(L-\alpha)^{-1},L].
\end{equation}
In our example, $P$ is, by construction, a second-order differential operator $P=p_2D^2+p_1D+p_0$. Equating in (\ref{LP}) the coefficients of $D^4$, $D^3$ and $D^2$ gives the relations
\[
 3p_{2,x}+p_x=0,\quad
 3p_{2,xx}+3p_{1,x}+q_x=0,\quad
 p_{2,xxx}+3p_{1,xx}+3p_{0,x}+up_{2,x}-2u_xp_2+up_x=0,
\]
from which $p_2,p_1$ and $p_0$ are easily expressed in terms of $p,q$ and $u$. The integration constants can be neglected, as before, since they correspond to the ambiguity of the definition of $p,q$ and the addition of a constant to the operator $P$ itself. The coefficients of $D$ and $D^0$ in (\ref{LP}) give exactly the same equations for $p$ and $q$ as in the calculation using $R$.

\begin{proposition}
Equations (\ref{Bsq.uvz}) and (\ref{Bsq.pqx}) are equivalent to equation (\ref{LP}) with $L=D^3+uD+v$ and
\begin{equation}\label{Bsq.P}
 P=-\frac{1}{3}pD^2 +\frac{1}{3}(p_x-q)D -\frac{1}{9}(2p_{xx}-3q_x+2up).
\end{equation}
\end{proposition}

Equations (\ref{Bsq.pqx}) define the nonlocal variables $p$ and $q$ as solutions of linear ODEs with respect to $x$. These equations should be supplemented with rules of differentiation with respect to $t_n$:
\begin{equation}\label{Bsq.pqtn}
 p_{t_n}=\partial_z(D^{-1}(u_{t_n})),\quad q_{t_n}=\partial_z(D^{-1}(v_{t_n}));
\end{equation}
for instance, the derivatives in virtue of the system (\ref{Bsq}) are given by
\begin{equation}\label{Bsq.pqt}
 p_t=-3p_{xx}+6q_x,\quad q_t=-2p_{xxx}+3q_{xx}-2up_x.
\end{equation}
The commutativity property $[\partial_{t_m},\partial_{t_n}]=0$ implies the operator equations
\begin{equation}\label{AA}
 A_{m,t_n}-A_{n,t_m}=[A_n,A_m].
\end{equation}
Since the negative symmetry is the generating function for $\partial_{t_n}$, this means that also $[\partial_z,\partial_{t_n}]=0$ and
\begin{equation}\label{PA}
 P_{t_n}-A_{n,z}(L-\alpha)=[A_n,P].
\end{equation}
Finally, the negative symmetries $\partial_{z_\alpha}$ corresponding to different values of the parameter are also commutative. Let $p^{(\alpha)}$ and $q^{(\alpha)}$ define a solution of equations (\ref{Bsq.pqx}) for a fixed value of parameter and let $P^{(\alpha)}$ be the operator (\ref{Bsq.P}) constructed by this solution, then such operators satisfy the relation
\begin{equation}\label{PP}
 P^{(\alpha)}_{z_\beta}(L-\beta)-P^{(\beta)}_{z_\alpha}(L-\alpha)=[P^{(\beta)},P^{(\alpha)}].
\end{equation}

To summarize, in this section we derived the negative symmetry of the Boussinesq hierarchy, linear in the fields $p$ and $q$, and obtained a Lax representation for it. In the next section we will look at an alternative method that allows us to replace (\ref{Bsq.pqx}) with lower order nonlinear equations.

\subsection{Squared eigenfunctions method}\label{s:Bsq.sq}

Equation (\ref{LP'}) is the compatibility condition for equations $L\psi=\lambda\psi$ and $\psi_z=P(L-\alpha)^{-1}\psi$, that is
\begin{gather}
\label{Bsq.psix}
 \psi_{xxx}=-u\psi_x+(\lambda-v)\psi,\\
\label{Bsq.psiz}
 \psi_z=\frac{1}{9(\lambda-\alpha)}\bigl(-3p\psi_{xx} +3(p_x-q)\psi_x-(2p_{xx}-3q_x+2up)\psi\bigr).
\end{gather}
Let us transform this to the matrix zero curvature representation
\begin{equation}\label{Uz}
 U_z=V_x+[V,U],
\end{equation}
by replacing (\ref{Bsq.psix}) and (\ref{Bsq.psiz}) with
\[
 \Psi_x=U\Psi,\quad \Psi_z=V\Psi,\quad 
  U=\begin{pmatrix}
     0 & 1 & 0 \\
     0 & 0 & 1 \\
    \lambda-v & -u & 0
 \end{pmatrix},\quad 
 \Psi=\begin{pmatrix}
  \psi\\ \psi_x\\ \psi_{xx}
 \end{pmatrix}.
\]
The first row of the matrix $V$ is read directly from the equation (\ref{Bsq.psiz}), and to calculate the remaining rows it is necessary to apply differentiation with respect to $x$ twice, excluding $\psi_{xxx}$ due to (\ref{Bsq.psix}). We will not need these rows explicitly, but it is easy to track down that $V$ is of the form
\begin{equation}\label{V}
 V=V_0+\frac{1}{9(\lambda-\alpha)}V_1
\end{equation}
where $V_0$ and $V_1$ do not depend on $\lambda$. Then (\ref{Uz}) implies the matrix Lax equation for $V_1$ 
\begin{equation}\label{V1x}
 V_{1,x}=[U(\alpha),V_1].
\end{equation}
Note that $\det(V_1-\mu I)$ gives two first integrals for the system (\ref{Bsq.pqx}), but they are very cumbersome and difficult to use for the order reduction. Instead, we will lower the order by comparing $V_1$ with a matrix constructed from solutions to the original and adjoint linear problems at $\lambda=\alpha$. Let $L\psi=\alpha\psi$ and $L^\dag\varphi=\alpha\varphi$, that is
\begin{equation}\label{Bsq.psiphix}
 \psi_{xxx}+u\psi_x+v\psi=\alpha\psi,\quad -\varphi_{xxx}-(u\varphi)_x+v\varphi=\alpha\varphi. 
\end{equation}
These solutions determine the column and row vectors
\[
 \Psi=(\psi,\, \psi_x,\, \psi_{xx})^t,\quad 
 \Phi=(\varphi_{xx}+u\varphi,\, -\varphi_x,\, \varphi),
\]
which solve the equations
\[
 \Psi_x=U(\alpha)\Psi,\quad \Phi_x=-\Phi U(\alpha).
\]
From here it follows that the matrix $\widetilde V_1=\Psi\Phi$ satisfies the equation (\ref{V1x}) and the scalar $\delta=\Phi\Psi$ is a first integral:
\begin{equation}\label{Bsq.delta}
 \delta=\varphi\psi_{xx}-\varphi_x\psi_x+\varphi_{xx}\psi+u\varphi\psi,\quad \delta_x=0.
\end{equation}
Now, let us introduce the variables
\begin{equation}\label{gw}
 g=\varphi\psi,\quad w=\varphi\psi_x-\varphi_x\psi.
\end{equation}
It is easy to see that all ratios $\partial^n_x(\psi)/\psi$ and $\partial^n_x(\varphi)/\varphi$, as well as any bilinear form of derivatives of $\varphi$ and $\psi$ can be expressed in terms of $g,w$ and their derivatives (and, of course, the original potentials $u$ and $v$). As a result, equations (\ref{Bsq.psiphix}) and (\ref{Bsq.delta}) turn into the coupled system of ODEs for $g$ and $w$:
\begin{equation}\label{Bsq.gwx}
\left\{\begin{aligned}
 g_{xx}&=\frac{3(g^2_x-w^2)}{4g}-ug+\delta,\quad \delta_x=0,\\
 w_{xx}&=\frac{g_xw_x}{2g}-\frac{w(g^2_x-w^2)}{8g^2}-\frac{\delta w}{2g}-\frac{u}{2}w+(2\alpha-2v+u_x)g.
\end{aligned}\right.
\end{equation}
The consistent evolution with respect to $t$ is determined by the equations $\psi_t=A\psi$ and $\varphi_t=-A\varphi$ where $A=A^\dag=3D^2+2u$, which imply
\begin{equation}\label{Bsq.gwt}
 g_t=3w_x,\quad
 w_t=-\frac{3g_x(g^2_x-w^2)}{8g^2}+\frac{3ww_x}{2g}-\frac{3\delta g_x}{2g}-\frac{3}{2}ug_x+u_xg,\quad
 \delta_t=0.
\end{equation}
Equations (\ref{Bsq.gwx}) and (\ref{Bsq.gwt}) determine an extension of the system (\ref{Bsq}) to nonlocal variables $w$ and $g$. Similarly, considering the equations $\psi_{t_n}=A_n\psi$ and $\varphi_{t_n}=-A^\dag_n\varphi$, one can derive formulas for the derivatives of $g$ and $w$ with respect to $t_n$.

It remains to compare the matrices $V_1$ and $c\widetilde V_1$, where $c$ is an arbitrary numerical coefficient chosen for convenience. To do this, it is enough to compare their first rows: 
\[
 (-2p_{xx}+3q_x-2up,\, 3p_x-3q,\, -3p)
\]
(according to (\ref{Bsq.psiz})) and
\[
 c\psi\Phi=c\psi(\varphi_{xx}+u\varphi,\, -\varphi_x,\, \varphi)
 =c\Bigl(\frac{2gg_{xx}-2gw_x-g^2_x+w^2}{4g}+ug,\, \frac{w-g_x}{2},\, g\Bigr).
\]
We set $c=-6$, then comparing the last two components of these vectors gives
\begin{equation}\label{Bsq.pqgw}
 p=2g,\quad q=g_x+w
\end{equation}
and it is easy to check, taking into account equations (\ref{Bsq.gwx}), that the first components differ only by the constant $2\delta$, which implies that $V_1=-6\widetilde V_1+2\delta I$. As a result, we arrive at the following statement, which can also be verified by direct calculation.

\begin{proposition}\label{pr:Bsq.uvzgw}
1) Equations (\ref{Bsq.gwx}) and (\ref{Bsq.gwt}) form a correct extension of the system (\ref{Bsq}) to the variables $w$ and $g$, that is, the derivations $\partial_x$ and $\partial_t$ determined by these formulas commute provided that $u$ and $v$ satisfy (\ref{Bsq}).

2) If $g$ and $w$ satisfy equations (\ref{Bsq.gwx}) and (\ref{Bsq.gwt}) then $p=2g$ and $q=g_x+w$ satisfy equations (\ref{Bsq.pqx}) and (\ref{Bsq.pqt}).

3) Equations
\begin{equation}\label{Bsq.uvzgw}
 u_z=2g_x,\quad v_z=g_{xx}+w_x
\end{equation}
define a negative symmetry for (\ref{Bsq}), in the sense that derivations $\partial_z$ and $\partial_t$ commute.
\end{proposition}

Equations (\ref{Bsq.gwx}) and (\ref{Bsq.uvzgw}) can also be viewed as an integrable system on its own, in the sense that they admit the zero curvature representation (\ref{Uz}). For completeness, we present the corresponding matrix $V$ (\ref{V}), constructed according to the scheme described above:
\begin{gather*}
 V=-\frac{1}{3}\begin{pmatrix}
  0 & 0 & 0\\
  2g & 0 & 0\\
  3g_x+w & 2g &0
 \end{pmatrix} +\frac{1}{6(\lambda-\alpha)g}AB,\\
 A^t=\left(2g,\, g_x+w,\, \frac{g^2_x-w^2}{4g}+w_x-ug+\delta\right),\quad
 B=\left(\frac{w^2-g^2_x}{4g}+w_x-ug-\delta,\, g_x-w,\, -2g\right).
\end{gather*}

Note that the change (\ref{Bsq.pqgw}) between $p,q$ and $g,w$ is invertible. However, it does not establish equivalence between systems (\ref{Bsq.pqx}) and (\ref{Bsq.gwx}), since they have different orders of derivatives: for the first system it is 8, and for the second only 4. The orders do not coincide even taking into account the first integral $\det(V_1-\mu I)$. Thus, the converse statement 2) of the Proposition \ref{pr:Bsq.uvzgw} is not true and the negative symmetry from the Proposition \ref{pr:Bsq.uvzpq} is more general. However, the equations (\ref{Bsq.gwx}) still define the generating function for higher symmetries. It is easy to show that they admit a formal solution in the form of series in $\zeta=\alpha^{-1/3}$: if we fix the value of the constant $\delta=3\zeta^3$ then 
\begin{equation}\label{gwseries}
\begin{aligned}
 g&= \zeta +\frac{u}{3\zeta} -\frac{u_x-2v}{3\zeta^2} -\frac{u_{xxx}-2v_{xx}+2uu_x-4uv}{9\zeta^4}\\
  &\qquad\qquad   -\frac{3u_{xxxx}+15uu_{xx}+45u_xv-45v^2+5u^3}{81\zeta^5} +\dotsc,\\
 w&= 2\zeta^2 -\frac{u_x-2v}{3\zeta} -\frac{u_{xx}+2u^2}{9\zeta^2} -\frac{3u_{xxxx}+18uu_{xx}+36u_xv-36v^2+8u^3}{81\zeta^4}+\dotsc
\end{aligned}
\end{equation}
where all coefficients are uniquely defined; moreover, the coefficients for all powers of $\zeta^{-3k}$ vanish. When substituting these series into (\ref{Bsq.uvzgw}), we obtain a generating function for the flows (\ref{Bsq.LAn}) with fixed numerical coefficients, in contrast to the formula (\ref{Bsq.Dz}) which contains arbitrary constants.

\subsection{Sawada--Kotera and Kaup--Kupershmidt reductions}\label{s:SKKK}

The hierarchy (\ref{Bsq.LAn}) admits two reductions to one-field hierarchies with simplest equation of fifth order. For $v=0$ (or $v=u_x$, which is equivalent under the change $L=D^3+uD$ by $-L^\dag$), the flow $\partial_{t_5}$ turns, up to a numerical factor, to the Sawada--Kotera equation \cite{Sawada_Kotera_1974} 
\begin{equation}\label{SK}
 u_{t_5}=u_{5x}+5uu_{xxx}+5u_xu_{xx}+5u^2u_x,
\end{equation}
and if $2v=u_x$ ($L=-L^\dag=D^3+uD+\frac{1}{2}u_x$) then the Kaup--Kupershmidt equation appears \cite{Kaup_1980} 
\begin{equation}\label{KK}
 u_{t_5}=u_{5x}+5uu_{xxx}+\frac{25}{2}u_xu_{xx}+5u^2u_x.
\end{equation}
In both cases, the formula (\ref{Bsq.LAn}) for the operators $A_n$ remains the same, but with additional restriction: $n\ne3k$ and $n\ne2k$, since the flows with even $n$ are not consistent with the reduction and should be rejected. For this reason, the recursion operators for these equations are of order 6 (explicit formulas can be found, for example, in \cite{Gurses_Karasu_Sokolov_1999, Wang_2002}). The equation (\ref{LAB}) is replaced with $L_{t_{n+6}}=L_{t_n}L^2+[B_n,L]$  and, consequently, equation (\ref{LP}) is replaced with
\begin{equation}\label{L2P}
 L_z(L^2-\alpha^2)=[P,L]  
\end{equation}
where $P$ is a differential operator of order not greater than 5. From here it is clear without calculations that the passage to the matrix representation (\ref{Uz}) brings to a matrix $V$ with poles at $\lambda=\pm\alpha$. The matrix coefficients at these poles satisfy the Lax equations (\ref{V1x}) with the matrices $U(\pm\alpha)$, which brings to a pair of coupled systems of the form (\ref{Bsq.gwx}). The generating function for the symmetries of equations (\ref{SK}) or (\ref{KK}) is given by the series (containing only powers $\zeta^{6k\pm1}$)
\[
 u_z=g_x(\zeta)-g_x(-\zeta)
\]
where $g(\zeta)$ is the series from (\ref{gwseries}), under the corresponding substitution $v=0$ or $2v=u_x$. The relation (\ref{Bsq.uvzgw}) for the second component takes the form
\[
 2v_z=g_{xx}(\zeta)-g_{xx}(-\zeta)+w_x(\zeta)-w_x(-\zeta)
\]
and is compatible with this substitution. For the case of reduction $2v=u_x$, equations (\ref{Bsq.gwx}) turns out to be invariant under the change $(\alpha,\delta,g,w)\leftrightarrow(-\alpha,-\delta,-g,w)$, which leads to the fact that the series $g$ contains only odd powers of $\zeta$ and $w$ contains only even powers. Due to this, the negative symmetry for the equation (\ref{KK}) can still be specified by the formula $u_z=2g_x$.

The system (\ref{Bsq.gwx}) should not be viewed only as a tool for constructing generating functions; its solutions can also be considered for a fixed parameter $\alpha$, including $\alpha=0$, although this does not make sense from the point of view of expansion (\ref{gwseries}). For equations (\ref{SK}) and (\ref{KK}), the value $\alpha=0$ is distinguished since
in this case it is possible to further reduce the dimension of the system. This leads to {\em degenerate negative symmetries}, which are described by simpler equations and are of independent interest. In the case of $v=0$, $\alpha=0$, equations (\ref{Bsq.uvzgw}) imply $g_x+w=c=\operatorname{const}$. Substituting this into the system (\ref{Bsq.gwx}), one finds that $c=0$ and the second equation of the system becomes a consequence of the first one. Similarly, in the case of $2v=u_x$, $\alpha=0$, it can be shown that $w=0$ and the second equation (\ref{Bsq.gwx}) is satisfied identically. Using the formula (\ref{Bsq.pqtn}) to determine the evolution with respect to $t_5$, we arrive at the following answers.

\begin{proposition}\label{pr:SKKK.neg0}
The Sawada--Kotera equation (\ref{SK}) admits the degenerate negative symmetry
\[
 u_z=g_x,\quad g_{xx}=-ug+\delta,\quad g_{t_5}=(2u_{xx}+u^2)g_x-(u_{xxx}+uu_x)g-3\delta u_x;
\]
the Kaup--Kupershmidt equation (\ref{KK}) admits the degenerate negative symmetry
\[
 u_z=g_x,\quad g_{xx}=\frac{3g^2_x}{4g}-ug+\delta,\quad g_{t_5}=(\tfrac{1}{2}u_{xx}+u^2)g_x-(u_{xxx}+4uu_x)g.
\]
In both cases, equations for $g$ define a correct extension of the evolution equation for $u$, and the equality
$(u_{t_5})_z=(u_z)_{t_5}$ holds.
\end{proposition}

\begin{remark}
The above equations allow the introducing of the potential $u=a_x$, $g=a_z$. For example, the potential forms of the Kaup--Kupershmidt equation and its degenerate negative symmetry are
\[
 a_{t_5}=a_{5x}+5a_xa_{xxx}+\frac{15}{4}a^2_{xx}+\frac{5}{3}a^3_x,\quad 
 a_{xxz}=\frac{3a^2_{xz}}{4a_z}-a_xa_z+\delta.
\]
The latter equation is reduced by additional point transformations to the Degasperis--Procesi equation \cite{Degasperis_Procesi_1999, Degasperis_Hone_Holm_2002}. Moreover, if $\delta=0$ then further degeneration is possible: the equation acquires the first integral \cite{Adler_Shabat_2012} 
\[
 a^{-3/2}_z(a_za_{xzz}-a_{xz}a_{zz})+\frac{2}{3}a^{3/2}_z=\gamma=\operatorname{const}
\]
and is reduced to the Tzitzeica equation $b_{xz}=\gamma e^{-b/2}-\frac{2}{3}e^b$ for $b=\log a_z$. 
\end{remark}

\section{Drinfeld--Sokolov system}\label{s:DS3}

\subsection{Lax representations and the recursion operator}\label{s:DS3.LR}

Our second example is related to the Drinfeld--Sokolov system DS-III \cite{Drinfeld_Sokolov_1985}
\begin{equation}\label{DS3}
 u_t= u_{xxx}-6uu_x-6v_x,\quad
 v_t= -2v_{xxx}+6uv_x.
\end{equation}
Most of the formulas for it turn out to be more complicated than for the Boussinesq equation, but the general scheme remains the same. The hierarchy of the system (\ref{DS3}) is determined by the Lax equations
\begin{equation}\label{DS3.LAn}
 L_{t_n}=[A_n,L],\quad A_n=(L^{n/4})_+,\quad n=1,3,5,\dots
\end{equation}
where $L$ is the general self-adjoint operator of fourth order
\begin{equation}\label{DS3.L}
 L=L^\dag=(D^2-u)^2+v=D^4-2uD^2-2u_xD-u_{xx}+u^2+v.
\end{equation}
The flow $\partial_t$ coincides with $4\partial_{t_3}$ and corresponds to the operator
\begin{equation}\label{DS3.A}
 A=4A_3=4(L^{3/4})_+=4D^3-6uD-3u_x.
\end{equation}

\begin{remark}
The same system arises with a different choice of operators:
\begin{equation}\label{DS3.LA'}
\begin{gathered}
 \tilde L=\tilde L^\dag= D^4-4uD^2-6u_xD-2u_{xx}-4v-2v_xD^{-1},\\
 \tilde A=-2(\tilde L^{3/4})_+ = -2D^3+6uD+6u_x.
\end{gathered}
\end{equation}
The relationship between both representations is explained in the next section.
\end{remark}

The method described in section \ref{s:Bsq.LR} (applied to any of the operators $L$ or $\tilde L$) brings to the recursion operator \cite{Gurses_Karasu_1999, Gurses_Karasu_Sokolov_1999}
\begin{equation}\label{DS3.R}
\begin{gathered}[b]
 16R=\begin{pmatrix}
  D^4-8uD^2-12u_xD-8u_{xx}+16u^2+16v & -10D^2+8u\\
  10v_xD+12v_{xx} & -4D^4+16uD^2+8u_xD+16v
 \end{pmatrix}\\
 -2\binom{u_{xxx}-6uu_x-6v_x}{-2v_{xxx}+6uv_x}D^{-1}(1,0) +4\binom{u_x}{v_x}D^{-1}(u,1).
\end{gathered}
\end{equation}
The operator $R$ sends the flow $\partial_{t_n}$ to the flow $\partial_{t_{n+4}}$ (coefficient 16 is chosen so that it corresponds to the normalization of operators $A_n$ adopted in (\ref{DS3.LAn})). The column factors at the integral terms in $R$ correspond to the flows $\partial_t=4\partial_{t_3}$ and $\partial_x=\partial_{t_1}$, which also play the role of the seed flows. The entire hierarchy consists of the flows $\partial_{t_{4k+1}}=R^k(\partial_{t_1})$ and $\partial_{t_{4k+3}}=R^k(\partial_{t_3})$. The explicit formula (\ref{DS3.LAn}) implies that the integral terms in $R$ do not produce nonlocalities, which means that any flow of the hierarchy can be represented as
\begin{equation}\label{DS3.utau}
 u_\tau=p_x,\quad uu_\tau+v_\tau=q_x
\end{equation}
where $p$ and $q$ are local functions of $u$, $v$ and their derivatives; for instance,  
\[
 4u_{t_3}= (u_{xx}-3u^2-6v)_x,\quad
 4uu_{t_3}+4v_{t_3}= \bigl(-2v_{xx}+uu_{xx}-\tfrac{1}{2}u^2_x-2u^3\bigr)_x.
\]
We define the negative flow $\partial_z$ as the generating function
\[
 \partial_z=(c_1\partial_{t_1}+c_2\partial_{t_3}) +\alpha^{-1}(c_1\partial_{t_5}+c_2\partial_{t_7}) +\alpha^{-2}(c_1\partial_{t_9}+c_2\partial_{t_{11}})+\dots
\]
which is also searched in the form (\ref{DS3.utau}). As before, the Lax equation (\ref{LP}) $L_z(L-\alpha)=[P,L]$ is derived, where $P$ is a third order differential operator. As a result of direct calculations, we obtain the following equations of negative symmetry:
\begin{gather}
\label{DS3.uvz}
 u_z=p_x,\quad v_z=q_x-up_x,\\[4pt]
\label{DS3.pqx}
\left\{\begin{aligned}
  & p_{5x}+2up_{3x}+8u_xp_{xx}+2(u_{xx}+4u^2+8(v-\alpha))p_x\\
  &\qquad -2(u_{3x}-6uu_x-6v_x)p -10q_{3x}+8uq_x+4u_xq=0,\\[2pt]
  & up_{5x} +4u_xp_{4x} +2(3u_{xx}-2u^2)p_{3x} +\bigl(4u_{3x}-10uu_x+\tfrac{5}{2}v_x\bigr)p_{xx}\\
  &\qquad +(u_{4x}-4uu_{xx}-2u^2_x+3v_{xx}-4u(v-\alpha))p_x +(v_{3x}-3uv_x)p\\
  &\qquad -q_{5x}+4uq_{3x}+2u_xq_{xx}+4(v-\alpha)q_x+v_xq=0.
\end{aligned}\right.
\end{gather}

\begin{proposition}
Equations (\ref{DS3.uvz}) and (\ref{DS3.pqx}) are equivalent to equation (\ref{LP}) for the operators $L$ (\ref{DS3.L}) and 
\begin{equation}\label{DS3.P}
 16P=8pD^3 -4p_xD^2 +2(p_{xx}-6up-2q)D-p_{xxx}-6u_xp+2up_x+6q_x,
\end{equation}
and also to the equation $R\binom{u_z}{v_z}=\alpha\binom{u_z}{v_z}$ with the operator (\ref{DS3.R}).
\end{proposition}

\subsection{Squared eigenfunctions method}\label{s:DS3.sq}

Equation (\ref{LP}) serves as the compatibility condition for the equations $L\psi=\lambda\psi$ and $\psi_z=P(L-\alpha)^{-1}\psi$, that is
\begin{gather}
\label{DS3.psix}
 \psi_{xxxx}=2u\psi_{xx}+2u_x\psi_x+(u_{xx}-u^2-v+\lambda)\psi,\\
\label{DS3.psiz}
 \psi_z=\frac{1}{16(\lambda-\alpha)}\bigl(
    8p\psi_{xxx} -4p_x\psi_{xx} +2(p_{xx}-6up-2q)\psi_x-(p_{xxx}+6u_xp-2up_x-6q_x)\psi\bigr).
\end{gather}
In order to obtain the zero curvature representation (\ref{Uz}), we write (\ref{DS3.psix}) and (\ref{DS3.psiz}) in the form
\[
 \Psi_x=U\Psi,\quad \Psi_z=V\Psi,\quad 
 U=\begin{pmatrix}
  0 & 1 & 0 & 0\\
  0 & 0 & 1 & 0\\
  0 & 0 & 0 & 1\\
  u_{xx}-u^2-v+\lambda & 2u_x & 2u & 0\\
 \end{pmatrix},\quad 
 \Psi=\begin{pmatrix}
  \psi\\ \psi_x\\ \psi_{xx}\\ \psi_{xxx} 
 \end{pmatrix}.
\]
As before, we need only the first row of the matrix $V$, which is determined intermediately by equation (\ref{DS3.psiz}). The dependence of $V$ on $\lambda$ is given by the formula
\[
 V=V_0+\frac{1}{16(\lambda-\alpha)}V_1
\]
and (\ref{Uz}) implies the Lax equation $V_{1,x}=[U(\alpha),V_1]$. Its particular solutions are constructed by formulas
\[
 \widetilde V_1=\Psi\Phi,\quad \Psi_x=U(\alpha)\Psi,\quad \Phi_x=-\Phi U(\alpha)
\]  
where, due to the self-adjointness of the operator $L$, both vectors $\Psi$ and $\Phi$ are determined by solutions $\psi$ and $\varphi$ of the same equation (\ref{DS3.psix}):
\[
 \Psi=\Psi[\psi]=(\psi,\, \psi_x,\, \psi_{xx},\, \psi_{xxx})^t,\quad
 \Phi=\Phi[\varphi]=(-\varphi_{xxx}+2u\varphi_x,\, \varphi_{xx}-2u\varphi,\, -\varphi_x,\, \varphi).
\]
The inner product $\delta=\Phi\Psi$ provides the first integral
\[
 \delta=\varphi\psi_{xxx}-\varphi_x\psi_{xx}+\varphi_{xx}\psi_x-\varphi_{xxx}\psi-2u(\varphi\psi_x-\varphi_x\psi),\quad 
 \delta_x=0.
\]
The variables $g=\varphi\psi$ and $w=\varphi\psi_x-\varphi_x\psi$ satisfy the following ODEs with respect to $x$:
\begin{equation}\label{DS3.gwx}
\left\{\begin{aligned}
 & 2ww_{xxxx}-2w_xw_{xxx}+w^2_{xx}=4u(2ww_{xx}-w^2_x)+4u_xww_x+4(v-\alpha)w^2+\delta^2,\\
 & wg_{xx}-w_xg_x+w_{xx}g=\frac{w}{2g}(g^2_x-w^2)+(2uw+\delta)g,\quad \delta_x=0.
\end{aligned}\right.
\end{equation}
The consistent $t$-evolution is defined by equations $\psi_t=A\psi$ and $\varphi_t=A\varphi$ with the operator (\ref{DS3.A}), which gives
\begin{equation}\label{DS3.gwt}
 w_t=-2w_{xxx}+6uw_x,\qquad g_t=\left(4g_{xx}-6ug-\frac{3(g^2_x-w^2)}{g}\right)_x,\quad \delta_t=0.
\end{equation}

\begin{remark}\label{rem:LA'}
Compared to the system (\ref{Bsq.gwx}), we see the peculiar property that equations for $w$ are separated (which is due to the self-adjointness of $L$). Moreover, the first equation (\ref{DS3.gwx}) serves as the first integral for the linear equation
\[
 w_{5x}-4uw_{3x}-6u_xw_{xx}-(2u_{xx}+4v-4\alpha)w_x-2v_xw=0,
\]
which brings to the second Lax repesentation for (\ref{DS3}) with operators (\ref{DS3.LA'}): indeed, $w_x$ satisfies the equation $\tilde Lw_x=4\alpha w_x$ and the equation for $w_t$ takes the form $w_{xt}=\tilde Aw_x$ after differentiation. 
\end{remark}

On the next step of our algorithm, we have to compare the matrices $V_1$ and $\widetilde V_1$. According to (\ref{DS3.psiz}), the first row of $V_1$ reads
\begin{equation}\label{DS3.V1-1}
 (-p_{xxx}-6u_xp+2up_x+6q_x,\, 2p_{xx}-12up-4q,\, -4p_x,\, 8p),
\end{equation}
and the first row of $\widetilde V_1=\Psi\Phi$ is
\[
 \psi(-\varphi_{xxx}+2u\varphi_x,\, \varphi_{xx}-2u\varphi,\, -\varphi_x,\, \varphi).
\]
From a comparison of the last two components it is clear that these formulas are not consistent. However, to obtain a coincidence it is enough to apply symmetrization: let us take $\widetilde V_1=8\Psi[\psi]\Phi[\varphi]+8\Psi[\varphi]\Phi[\psi]$, then the first row of this matrix coincides with (\ref{DS3.V1-1}) where
\[
 p=2\varphi\psi,\quad q=2u\varphi\psi-\varphi\psi_{xx}+2\varphi_x\psi_x-\varphi_{xx}\psi.
\]
By passing to the variables $g$ and $w$, we obtain the desired change
\begin{equation}\label{DS3.pqgw}
 p=2g,\quad q=-g_{xx}+\frac{g^2_x-w^2}{g}+2ug
\end{equation}
and define the derivation $\partial_z$ according to the formula (\ref{DS3.uvz}). This gives
\begin{equation}\label{DS3.uvzgw}
 u_z= 2g_x,\quad v_z= -g_{xxx}+\left(\frac{g^2_x-w^2}{g}\right)_x+2u_xg
\end{equation}
and, finally, we arrive at the following result.

\begin{proposition}\label{pr:DS3z}
1) Equations (\ref{DS3.gwx}) and (\ref{DS3.gwt}) are consistent, provided that $u$ and $v$ satisfy (\ref{DS3}), that is, these equations define a correct extension of the system (\ref{DS3}) to the variables $w$ and $g$.

2) If $g$ and $w$ satisfy (\ref{DS3.gwx}), then $p$ and $q$ defined by equalities (\ref{DS3.pqgw}) satisfy (\ref{DS3.pqx}).

3) The derivation (\ref{DS3.uvzgw}) commutes with (\ref{DS3}). 
\end{proposition}

Similar to the Boussinesq equation example, the change (\ref{DS3.pqgw}) reduces more general equations for $p$ and $q$ to simpler equations for $g$ and $w$. However, the system (\ref{DS3.gwx}) remains general enough and still admits solutions in the form of formal power expansions in $\alpha^{-1/2}$. In particular, the formal solution of the first equation (\ref{DS3.gwx}) is given by series
\begin{equation}\label{wseries}
 w= w_0+\frac{w_1}{(-4\alpha)}+\frac{w_2}{(-4\alpha)^2}+\dotsc,\qquad
 \delta^2=4\alpha\Bigl(w^2_0+\frac{\delta_1}{(-4\alpha)}+\frac{\delta_2}{(-4\alpha)^2}+\dots\Bigr)
\end{equation}
where $\delta_n$ and $w_0\ne0$ are arbitrary constants and all remaining coefficients $w_n$ are uniquely calculated by the recurrent formula
\begin{gather*}
 2w_0w_{n+1} = \sum^n_{s=0}\bigl(
  2w_sw_{n-s,xxxx}-2w_{s,x}w_{n-s,xxx}+w_{s,xx}w_{n-s,xx} -4u(2w_sw_{n-s,xx}-w_{s,x}w_{n-s,x}) \\
  -4u_xw_sw_{n-s,x}-4vw_sw_{n-s}\bigr) -\sum^n_{s=1}w_sw_{n+1-s} +\delta_{n+1}.
\end{gather*}
By setting $w_0=-1/2$ and $\delta_1=0$ (without loss of generality, due to the change $v\to v+\operatorname{const}$), we find
\begin{gather*}
 w_1=v,\qquad w_2=v_{xxxx}-4uv_{xx}-2u_xv_x-3v^2_x-\delta_2,\\ 
 w_3= v_{8x} -8uv_{6x} -20u_xv_{5x} 
     -2(16u_{xx}-8u^2+5v)v_{4x}
     -2(14u_{xxx}-24uu_x+11v_x)v_{xxx}
     -19v^2_{xx}\\ \qquad
     -4(3u_{4x}-8uu_{xx}-3u^2_x-10uv)v_{xx}
     -2(u_{4x}-4uu_{xx}+u^2_x-10uv)_xv_x
     +10v^3+2\delta_2v-\delta_3
\end{gather*}
and so on. The second equation (\ref{DS3.gwx}) makes possible to construct the formal series for $g$, but since this equation includes $\delta$, and not $\delta^2$, the expansion is carried out in powers of $\alpha^{-1/2}$ rather than $\alpha^{-1}$.

\subsection{Krichever--Novikov equation}\label{s:KN}

The system (\ref{DS3}) admits the obvious reduction $v=0$, which brings to the KdV equation
\[
 u_t=u_{xxx}-6uu_x.
\]
Let us see how the formulas for the negative symmetry are simplified in this case. The first equation (\ref{DS3.gwx}) with $v=0$ admits a trivial solution $w=\operatorname{const}$, $4\alpha w^2=\delta^2$. Since the formal series for $w$ is constructed uniquely, this constant will be its value. Then the second equation (\ref{DS3.gwx}), equations (\ref{DS3.gwt}) and (\ref{DS3.uvzgw}) after simple transformations are reduced to
\[
 g_{xx}=\frac{g^2_x-w^2}{2g}+(2u+{\delta}{w})g,\quad g_t= g_{xxx}-6ug_x,\quad u_z=2g_x;
\]
the second equation (\ref{DS3.uvzgw}) turns into identity $v_z=0$. Up to notations, this coincides with equations (\ref{KdVneg}) from Introduction.

This reduction is not the only one possible. In fact, the system (\ref{DS3}) allows an infinite series of reductions defined by the polynomiality condition of the series (\ref{wseries}), that is, by equations $0=w_n=w_{n+1}=w_{n+2}=\dotso$. Indeed, since $w$ satisfies linear equations (see Remark \ref{rem:LA'}), it follows that such a termination defines an invariant submanifold consistent with the dynamics in $x$ and $t$. We obtain the KdV equation for $n=1$, and if $n=2$ then the Krichever--Novikov equation appears \cite{Krichever_Novikov_1980}
\begin{equation}\label{KN}
 v_t= v_{xxx}-\frac{3(v^2_{xx}-r(v))}{2v_x},\quad r(v)=-2v^3-2\delta_2v-\delta_3.
\end{equation}
As can be seen from the above expression for $w_3$, the next possible choice $n=3$ turns out to be too complicated and leads to a highly non-local equation, so we restrict ourselves to the case $n=2$.

In more detail, let
\begin{equation}\label{KNw}
 w=-\frac{1}{2}-\frac{v}{4\alpha},\quad 
 \delta^2= \alpha+\frac{\delta_2}{4\alpha}-\frac{\delta_3}{16\alpha^2}
  =\frac{r(-2\alpha)}{16\alpha^2}
\end{equation}
then one can prove that the first equation (\ref{DS3.gwx}) is equivalent to one relation
\begin{equation}\label{KNu}
 u=\frac{2v_xv_{xxx}-v^2_{xx}+r(v)}{4v^2_x}.
\end{equation}
This substitution is known for long \cite{Svinolupov_Sokolov_YamilovIbragimov_1983} and it defines the constraint which reduces the system (\ref{DS3}) to equation (\ref{KN}). Note that in the original paper \cite{Krichever_Novikov_1980}, this constraint was obtained from other considerations, namely from the commutativity condition $[L,M]=0$ with a sixth-order differential operator $M$.

The equations defining negative symmetry remain valid. In addition to the substitutions (\ref{KNu}) and (\ref{KNw}), we change the notation of the parameters $-2\alpha=\beta$ and $4\alpha\delta=\gamma$. Then equations (\ref{DS3.gwx}) and (\ref{DS3.gwt}) are reduced to the following equations for the nonlocal variable $g$ and its $t$-evolution:
\begin{equation}\label{KNg}
 2gg_{xx}-g^2_x-\frac{2v_xgg_x}{v-\beta}
  -\left(\frac{2v_{xxx}}{v_x}-\frac{v^2_{xx}-r(v)}{v^2_x}-\frac{2(v_{xx}+\gamma)}{v-\beta}\right)g^2
  +\frac{(v-\beta)^2}{4\beta^2}=0,
\end{equation}
\begin{equation}\label{KNgt}
\begin{aligned}[b]
 g_t&=\left(\frac{v_{xxxx}}{v_x}-\frac{2v_{xx}v_{xxx}}{v^2_x}
  +\frac{v^3_{xx}-r(v)v_{xx}}{v^3_x}
  -\frac{4v^2_{xx}+14v^3-6\beta v^2+10\delta_2v-2\delta_2\beta+4\delta_3}{2(v-\beta)v_x}\right)g\\ 
 &\qquad -\left(\frac{v_{xxx}}{v_x}-\frac{v^2_{xx}-r(v)}{2v^2_x}
  -\frac{2(v_{xx}-\gamma)}{v-\beta}\right)g_x,\quad \gamma^2=r(\beta),
\end{aligned}
\end{equation}
and equations (\ref{DS3.uvzgw}) for the negative flow $\partial_z$ turn into
\begin{equation}\label{vz}
 v_z= \frac{(v_xg_x-v_{xx}g)^2-r(v)g^2}{4(v-\beta)v_xg}+\frac{(v-\beta)v_x}{8\beta^2g}.
\end{equation}

\begin{proposition}\label{pr:KNz}
Equations (\ref{KNg}) and (\ref{KNgt}) define an extension of equation (\ref{KN}) to the variable $g$, that is, they are consistent provided that $v$ solves (\ref{KN}). The derivation (\ref{vz}) commutes with (\ref{KN}). 
\end{proposition}

\section{Concluding remarks}

This paper proposes a method for deriving negative symmetries based on the Lax representation in the algebra of differential operators. As shown in the paper \cite{Gurses_Karasu_Sokolov_1999}, the derivation scheme of the recursion operator also works for other types of Lax representations, for example, matrix ones, which suggests that our scheme for negative symmetries allows corresponding generalizations as well. Possible applications of negative symmetries, in addition to their role as generating functions, can be related with the construction of finite-dimensional reductions, including the Painlev\'e type ones, as shown in \cite{Adler_2023, Adler_2020, Adler_Kolesnikov_2023} for simpler examples related to the KdV equation and the Volterra lattice. Work in these directions will be continued in subsequent publications.

\subsubsection*{ACKNOWLEDGMENTS}
The reported study was funded by RFBR and SC RA, project number 20-52-05015.

\subsubsection*{AUTHOR DECLARATIONS}
\subsubsection*{Conflict of Interest}
The author has no conflicts to disclose.

\subsubsection*{DATA AVAILABILITY}
Data sharing is not applicable to this article as no new data were created or analyzed in this study.


\end{document}